\documentclass[aps,preprint,amsmath,amssymb]{revtex4-1}

\usepackage{graphicx, color}

\def\beq{\begin{equation}}
\def\eeq{\end{equation}}
\def\be{\begin{eqnarray}}
\def\ed{\end{eqnarray}}
\def\non{\nonumber}
\def\ga{\gamma}

\begin{document} 
\title{\Large \bf Direct CP Violation in 
Charm  Decays due to  Left-Right Mixing}

\date{\today}
  
\author{ \bf  Chuan-Hung Chen$^{1,2}$\footnote{Email:
physchen@mail.ncku.edu.tw}, Chao-Qiang Geng$^{3,2}$\footnote{Email: geng@phys.nthu.edu.tw} and Wei Wang$^4$\footnote{Email:wwang@ihep.ac.cn}
 }

\affiliation{ $^{1}$Department of Physics, National Cheng-Kung
University, Tainan 701, Taiwan \\
$^{2}$National Center for Theoretical Sciences, Hsinchu 300, Taiwan\\
$^3$ Department of Physics, National Tsing Hua University, Hsinchu 300, Taiwan\\
$^4$ Department of Physics, Huazhong University of Science and Technology, Wuhan 430074, China
 }

\begin{abstract}
Motivated by the $3.8\sigma$ deviation from no CP violation hypothesis for
the CP asymmetry (CPA) difference between $D^0\to K^+ K^-$ and $D^0\to \pi^+ \pi^-$,
 reported recently by LHCb and CDF,
 we investigate the CP violating effect due to the left-right (LR) mixing  in the general LR symmetric model. 
 In particular, in the non-manifest LR model we show that the large CPA difference could be explained, while the constraints from  $(\epsilon'/\epsilon)_K$ and $D^0$-$\bar D^0$ are satisfied.

\end{abstract}
\maketitle

In the standard model (SM),  we expect that the CP asymmetries (CPAs) in $D^0$ decays, defined by
\beq\label{CPA}
A_{CP}(D^{0}\to f) \equiv \frac{\Gamma(D^0\to f)-\Gamma(\bar D^0\to f)}
  {\Gamma(D^0\to f)+\Gamma(\bar D^0\to f)}\,,~~~(f=K^{+}K^{-}\,,~\pi^{+}\pi^{-})
\eeq
should be vanishingly small, and therefore an observation of a large CPA in the charm sector 
clearly indicates physics beyond the SM.
Recently, both LHCb~\cite{Aaij:2011in} and CDF~\cite{CDF} collaborations have seen a large
difference between the time-integrated CPAs in the
decays $D^0 \to K^+K^-$ and $D^0 \to \pi^+\pi^-$,
$\Delta A_{CP} \equiv A_{CP}(D^{0}\to K^{+}K^{-})-A_{CP}(D^{0}\to \pi^{+}\pi^{-})$, 
given by
\be\label{LHCbData}
\Delta A_{CP} &=&(-0.82\pm0.21(\text{stat.})\pm0.11(\text{sys.}))\%\,~~(\text{LHCb})\,,
\nonumber\\
&=& (-0.62 \pm 0.21(\text{stat.}) \pm 0.10(\text{sys.}))\%\,~~(\text{CDF})\,,
\ed
based on 0.62 fb$^{-1}$ and 9.7 fb$^{-1}$ of data, respectively.
By combing the above results with fully uncorrelated uncertainties, one obtains 
 the average value~\cite{CDF}
 \be
 \Delta A^{\rm avg}_{CP}& =& (-0.67 \pm 0.16 )\%\,,
  \label{NewData}
 \ed
which is about $3.8\sigma$ away from zero.

As the  time dependent CPA involves both direct  and indirect 
parts, $i.e.$ $A_{CP}^{dir}(D^{0}\to f)$ and  $A_{CP}^{ind}(D^{0}\to f)$,
one gets~\cite{Aaij:2011in} 
\be
\Delta A_{CP} &\simeq & \Delta A_{CP}^{dir}+ (9.8\pm0.3)\% A_{CP}^{ind}\,,  
\label{eq:DirCP}
\ed
where $\Delta A^{dir}_{CP}\equiv A_{CP}(D^{0}\to K^{+}K^{-})-A_{CP}(D^{0}\to \pi^{+}\pi^{-})$
and $A_{CP}^{ind}\equiv A_{CP}^{ind}(D^{0}\to f)$ is universal for $f=K^{+}K^{-}$ and $\pi^{+}\pi^{-}$
 and less than $0.3$\% due to the mixing parameters. 
It is clear that the average value in Eq.~(\ref{NewData})
is dominated by the difference of the direct CP asymmetries, $\Delta A_{CP}^{dir}$.

In order to have a nonzero direct CPA, two amplitudes $A_1$ and $A_2$ with both nontrivial  weak    and strong phase differences,
$\theta_W$ and $\delta_{S}$, are essential, giving the CPA
\be
\label{eq:DirCPA}
A_{CP}(D^0\to f) 
&=& {-2|A_1||A_2|\sin\theta_W\sin\delta_{S}\over |A_1|^2+|A_2|^2+2|A_1||A_2|\cos\theta_W\cos\delta_{S}}\,.
\ed
The SM description of   the direct CPA  for $D^0\to f$ arises from the interference between tree and penguin contributions, in which decay amplitudes have the generic expressions
\begin{eqnarray}
 A^q_{SM}(D^0\to f)= V_{cq}^* V_{uq}  \left( T^{q}_{SM}   + E^q_{SM} e^{i\delta^q_S}\right)- V_{cb}^* V_{ub} P^q_{\rm SM} e^{i \phi^q_S},
\end{eqnarray}
where $q=(d, s)$ represents  $f=(\pi^+ \pi^-, K^+K^-)$, respectively,  $V_{q'q}$  is the 
Cabibbo-Kobayashi-Maskawa (CKM) matrix element, $T'_{SM}(P'_{SM})$ denotes the tree (penguin) contribution in the SM,   
$E'_{SM}$ stands for the contributions of W-exchange topology, and $\delta^q_S (\phi^q_S)$ is the associated CP-even phase.
Due to  the hierarchy in the CKM matrix elements $V_{cq}^* V_{uq}\gg V_{cb}^* V_{ub}$,  the direct CPA could be estimated by
 \be
 A^{dir}_{CP}(D^0\to f) & \sim & Im\left( \frac{V^*_{cb} V_{ub}}{V^*_{cq} V_{uq}} \right)\frac{2 P^q_{SM}}{ |T^{q }_{SM}+E^q_{SM} e^{i\delta^q_S}|^2 }\left( T^q_{SM} \sin\phi^q_S + E^q_{SM} \sin(\delta^q_S - \phi^q_S) \right)  \,.
 \ed
 With 
${\rm Im}(V^*_{cb} V_{ub}/V^*_{cq} V_{uq})\approx \pm A^2 \lambda^4 \eta$,
$E^q_{SM}\sim T^q_{SM}$, and $\sin\phi^q_S\sim \sin(\delta^q_S- \phi^q_S)\sim O(1)$,  we could have  
 \be
 A_{CP}(K^- K^+) &\sim& -A_{CP}(\pi^- \pi^+)\sim -A^2 \lambda^4 \eta {P^q_{SM}\over T^q_{SM}} \,.
 \ed
Unless  $P^q_{SM}$ could be enhanced to several orders larger than $T^q_{SM}$ by some unknown QCD effects, normally the predicted $\Delta A_{CP}$ in the SM is far below the central value in Eq. (\ref{NewData}). The detailed analysis by various approaches in the SM can be referred to 
Refs.~\cite{Bigi:2011,Cheng:2012wr,Feldmann:2012js,Li:2012cf,Franco:2012ck}.
Clearly, a solution to the large $\Delta A_{CP}$ in Eq.~(\ref{NewData}) is
to introduce some new CP violating mechanism beyond the CKM.
 
 To understand the LHCb and CDF data, many theoretical 
 studies~\cite{Bigi:2011,Cheng:2012wr,Feldmann:2012js,Li:2012cf,Franco:2012ck,Isidori:2011,Brod:2011re,Wang:2011uu,Gersabeck:2011xj,arXiv:1111.6949,Hochberg:2011ru,Pirtskhalava:2011va,Chang:2012gn,Giudice:2012qq,Bhattacharya:2012ah,Altmannshofer:2012ur,Chen:2012am,Lodone:2012kp,Inguglia:2012ik,Brod:2012ud,Hiller:2012wf,Mannel:2012hb,Grossman:2012eb,Cheng:2012xb,Isidori:2012yx,KerenZur:2012fr,Barbieri:2012bh,HYCheng}
 have been  done.
  Since the mixing induced CPA in $D$-meson now is limited to be less than around $0.3\%$ and no significant evidence shows a non-vanishing CPA, if a large $\Delta A_{CP}$ indicates some new physics effects, the same mechanism contributing  to $A^{ind}_{CP}$ should be small or negligible. 
  To satisfy the criterion of a small $A^{ind}_{CP}$, it is interesting to explore the tree induced  new CP violating effects  in which the loop contributions are automatically suppressed. 
  In this paper, we examine the new CP source 
   associated with  right-handed charged currents and the left-right (LR) mixing angle, $\xi$, in a general $SU(2)_L\times SU(2)_R \times U(1)$ model \cite{Pati:1973rp,Langacker:1989xa}.  It is known that  the unitarity of the CKM matrix  gives a strict limit on  $\xi$ \cite{Wolfenstein:1984ay}.
    However, it was found that the allowed value of the mixing angle indeed could be as large as of order of $10^{-2}$ when the right-handed mixing matrix has 
    a different pattern from the CKM and carries large CP phases \cite{Langacker:1989xa}. The constraints from rare $B$ decays could be referred to Refs.~\cite{Grzadkowski:2008mf,Crivellin:2011ba}. Based on the possible large new CP phases and sizable $\xi$, we study the impact on the direct CPAs in $D^0\to f$ decays.

In terms of the notations in Ref.~\cite{Langacker:1989xa}, we  first write the mass eigenstates of charged gauge bosons as
\be
   \left( \begin{array}{c}
  W^\pm_L  \\ 
    W^\pm_R 
  \end{array} \right)=
  \left( \begin{array}{cc}
    \cos\xi &-\sin\xi \\ 
    e^{i\omega}\sin\xi & e^{i\omega} \cos\xi \\ 
  \end{array} \right) \left( \begin{array}{c}
  W^\pm_1  \\ 
    W^\pm_2
  \end{array} \right)\,.
 \ed
The phase $\omega$ arises from the complex vacuum expectation values (VEVs) of bidoublet scalars which are introduced to generate  the masses of fermions.  Since $m_W\ll m_{W_R}$,  it is more useful to take the approximation of $\cos\xi\approx 1$ and $\sin\xi\approx \xi$ . 
Accordingly, the charged current interactions in the flavor space can be expressed by 
 \be
 -{\cal L}_{CC} &=& \frac{1}{\sqrt{2}} \bar U \ga_\mu \left( g_L V^L P_L + g_R \xi \bar V^R P_R \right) D W^+_1 \non \\ &+&  \frac{1}{\sqrt{2} } \bar U \ga_\mu \left( -g_L \xi V^L P_L + g_R \bar V^R P_R\right) D W^+_2 +h.c.
 \ed 
where the flavor indices are suppressed, $V^L$ is the  CKM matrix, $\bar V^R=e^{i\omega}V^R$ and $V^R$ is the flavor mixing matrix for right-handed currents.  Consequently, the four-Femi interactions for $c\to u q\bar q$ induced by the LR mixing are  given by 
 \be
 {\cal H}^q_{\chi \chi'}&=& \frac{4G_F}{\sqrt{2}} \frac{g_R}{g_L} \xi \left[ V^{\chi'}_{uq} V^{\chi^*}_{cq}  \left( C'_1(\mu) (\bar u q)_{\chi'} (\bar q c)_{\chi}) + C'_2(\mu) (\bar u_\alpha q_\beta)_{\chi'}  (\bar q_\beta c_\alpha)_\chi \right) \right. \non \\
&+&  \left. V^{\chi}_{uq} V^{\chi'^*}_{cq}  \left( C'_1(\mu)  (\bar u q)_{\chi} (\bar q c)_{\chi'}) + C'_2(\mu)  (\bar u_\alpha q_\beta)_{\chi}  (\bar q_\beta c_\alpha)_{\chi'}  \right)  \right]\,,
 \ed
where $\chi=L(R)$ and $\chi'=R(L)$ while $q=s(d)$, and $(\bar q q')_{L(R)}=\bar q \gamma^\mu P_{L(R)}q'$. The Wilson coefficients $C'_{1}=\eta_{+}$ and $C'_{2}=-(\eta_+ -\eta_{-})/3$ with QCD corrections could be estimated by \cite{Cho:1993zb,Chang:1994wk}
  \be
   \eta_{+}  &=&  \left(\frac{\alpha_s(\mu)}{\alpha_s(m_c)}\right)^{-3/27}\left( \frac{\alpha_s(m_c)}{\alpha_s(m_b)}\right)^{-3/25} \left(\frac{\alpha_s(m_b)}{\alpha_s(m_W)}\right)^{-3/23}\,, \non \\
   \eta_{-} &=& \eta_{+}^{-8}\,.
  \ed
Due to the suppression of $g^2_R/m^2_R$, as usual we neglect the $W_R$ itself contributions \cite{Langacker:1989xa,Chang:1994wk}.   

Based on the decay constants and transition form factors, defined by
 \be
\langle 0 | \bar q' \ga^\mu \ga_5 q|P(p)\rangle &=& if_P p^\mu\,, \non \\
 \langle P(p_2)| \bar q \ga_\mu c| D (p_1)\rangle &=&
F^{DP}_{+}(k^2)\Big\{Q_{\mu}-\frac{Q\cdot k }{k^2}k_{\mu} \Big\} \non \\
&+&\frac{Q\cdot k}{k^2}F^{DP}_{0}(k^2)\,k_{\mu}\,,
 \ed
respectively,
with $Q=p_1+p_2$ and $k=p_1-p_2$, the decay amplitude for $D^0\to f$  in the QCD factorization approach is found to be
 \be
 %
 %
 A^q_{LR}(D^0\to f) &=& \left( \bar V^{R^*}_{cq} V^L_{uq} -V^{L^*}_{cq} \bar V^R_{uq} \right)T^q_{L R} \label{eq:AmpLR}
 \ed
 with
  \be
 T^d_{LR}&=& \frac{G_F}{\sqrt{2}} \frac{g_R}{g_L} \xi a'_1 f_\pi F^{D\pi}_0 ( m^2_D -m^2_\pi)\,, \non \\
 T^s_{LR}&=& \frac{f_K}{f_\pi} \frac{F^{DK}_0}{F^{D\pi}_0}\frac{m^2_D - m^2_K}{m^2_D -m^2_\pi}T^d_{RL}\,, \non
 \ed
 and $a'_1 = C'_1 + C'_{2}/N_c $. The associated branching ratio could be obtained by ${\cal B}(D^0\to f)= \tau_D |\vec{p_f}|A^q(D^0\to f)|^2/8\pi m_D^2$, where $\tau_D$ is the lifetime of the $D^0$ meson, $|p_{f}|$ is the magnitude of the $\pi (K)$ momentum and $A^q=A^q_{SM}+A^q_{LR}$.  With $V^L_{us}\approx -V^L_{cd}\approx \lambda$, the squared amplitude differences between $D^0\to f$ and its CP conjugate are 
\be
|A^d|^2 -|\bar A^d|^2 &=& -4 E^d_{SM}T^d_{LR}\sin\delta^d_S \frac{a_1'}{a_1} \xi \left( \lambda^2 Im V^{R}_{ud} - \lambda Im V^R_{cd} \right)\,, \non \\
|A^s|^2 -|\bar A^s|^2 &=& -4 E^s_{SM}T^s_{LR}\sin\delta^s_S  \frac{a_1'}{a_1} \xi  \left( \lambda Im V^{R}_{us} + \lambda^2 Im V^R_{cs} \right)\,. \label{eq:damp2}
\ed
Clearly, the direct CPA in $D^0\to f$ decay will strongly depend on the CP violating phases in $V^R_{cq, uq}$. 
 Since the (pseudo) manifest   LR model, denoted by $V^L=V^{R^{(*)}}$, has a strict limit on $\xi$, in this paper, we only focus on the non-manifest LR model, where except the unitarity, the elements in $V^R$ are arbitrary free parameters.

\begin{figure}[htbp]
\includegraphics*[width=5 in]{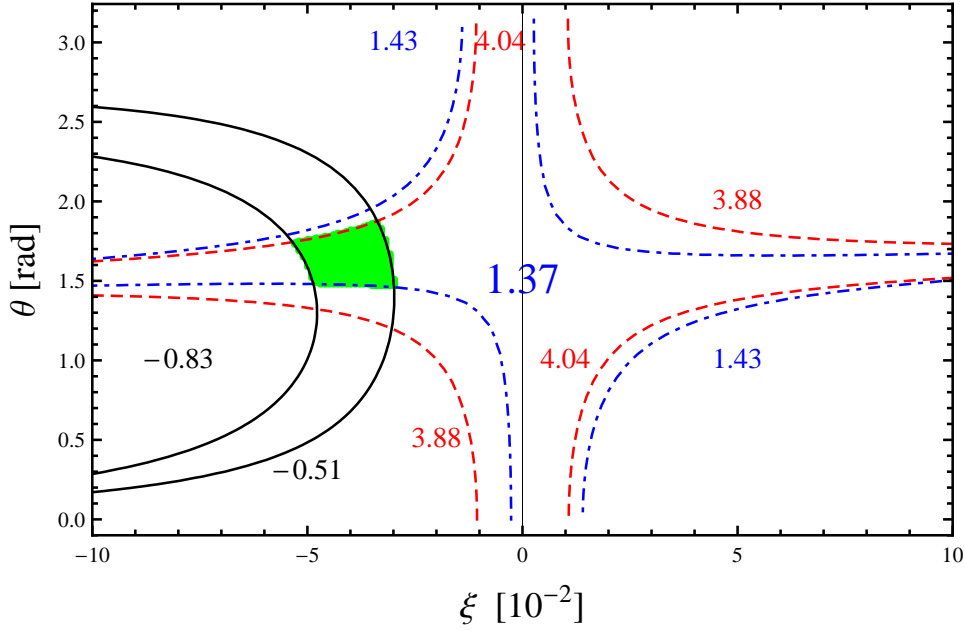}
\caption{BRs for  $D^0\to \pi^+ \pi^-$ (dashed) and $D^0\to K^+ K^-$ (dash-dotted) and $\Delta A_{CP}$ (solid), where the shaded band represents the allowed region. }
 \label{fig:A}
\end{figure}

In the numerical calculations, the input values of the SM are listed in Table~\ref{tab:inputs} \cite{Cheng:2012wr,Cheng:2010ry,PDG2010},
 where the resulting branching ratios (BRs) for $D^0\to (\pi^- \pi^+, K^-K^+)$ are estimated as $(1.38, 3.96) \times 10^{-3}$, while the current data are ${\cal B}(D^0\to \pi^- \pi^+ )=(1.400 \pm 0.026)\times 10^{-3}$ and ${\cal B}(D^0\to K^- K^+)= (3.96\pm 0.08)\times 10^{-3}$ \cite{PDG2010}.  
\begin{table}[hptb]
\caption{Numerical inputs for the parameters in the SM.
 } \label{tab:inputs}
\begin{ruledtabular}
\begin{tabular}{cccccc}
  $T^d_{SM}$& $T^s_{SM}$& $E^d_{SM}$ & $E^s_{SM}$ & $\delta^d_S$ & $\delta^s_S$
 \\ \hline
 $3.0\times 10^{-6}$GeV & $4.0\times 10^{-6}$GeV & $1.3\times 10^{-6}
 $GeV & $1.6\times 10^{-6}$GeV & $145^{\circ}$ & $108^{\circ}$ \\ \hline
 $V^L_{us}$ & $m_{\pi(K)}$ & $m_D$ & $f_{\pi(K)}$ & $F^{D\pi(K)}_0$& $m_t$   \\ \hline
 $0.22$  & $0.139 (0.497)$GeV & $1.863$GeV & $0.13(0.16)$GeV &
 $0.666(0.739)$ & 162.8 GeV  \\ 

\end{tabular}
\end{ruledtabular}
\end{table}
Although the QCD related  SM inputs  are extracted from the Cabibbo allowed decays, the influence of the new effects  on these decays is small. Due to 
the W-exchange topology  dominated by the final state interactions, the short-distance effects could be ignored. 
It is known that  the box diagrams with $W_L$ and $W_R$ yield  important contributions to
the $K^0$-$\bar K^0$ mixing \cite{Beall:1981ze}. 
However, due to the quarks in the diagrams for the D-system being down-type ones, we find the enhancement on the
$D^0$-$\bar D^0$ oscillation is small. Hence, the constraint from $\Delta m_D$ could be ignored.  
Since the CPAs involve $V^R_{ud, us}$,  we need to consider the constraint from the direct CPA in $K\to \pi \pi $ decays. Using the result in Ref.~\cite{He:1988th}, we know $(\epsilon'/\epsilon)_K \sim 1.25\times 10^{-3}g_R/g_L \xi  Im(\bar V^R_{us} - \lambda \bar V^{R^*}_{ud} )$. 
 Therefore, to avoid the constraint from $(\epsilon'/\epsilon)_K$, we adopt two cases: 
 (I) $Im(V^R_{us,ud})\to 0$  and (II) $Im(\bar V^R_{us}) \approx \lambda Im(\bar V^{R^*}_{ud} )$ \cite{He:1988th}.  
 We investigate the two cases separately as follows:

\underline{\it{Case} I}: In this case, Eq.~(\ref{eq:damp2}) is simplified as
 \be
|A^d|^2 -|\bar A^d|^2 &=& 4 E^d_{SM}T^d_{LR}\sin\delta^d_S \frac{a_1'}{a_1} \xi  \lambda Im \bar V^R_{cd} \,, \non \\
|A^s|^2 -|\bar A^s|^2 &=& -4 E^s_{SM}T^s_{LR}\sin\delta^s_S  \frac{a_1'}{a_1} \xi\lambda^2 Im \bar V^R_{cs}\,. \label{eq:damp2-1}
\ed
In general,  $V^R_{cd}$ and $V^R_{cs}$ are free parameters. In order to illustrate the impact of the LR mixing effects on $\Delta A_{CP}$ 
and make the CPAs of $\pi^+  \pi^-$ and $K^+ K^-$ modes to be more correlated, an interesting choice is 
$\bar V^R_{cd} \approx -\lambda e^{i\theta} $ and $\bar V^R_{cs} \approx e^{i\theta}$.  
Hence, the involving free parameters for the CPAs are the CP phase $\theta$ and the mixing angle $\xi$.  
Using Eqs.~(\ref{eq:AmpLR}) and $A^q=A^q_{SM}+A^q_{LR}$,   BRs for $D^0\to \pi^+ \pi^-$ (dashed) and $D^0\to K^+ K^-$ (dash-dotted)  as
functions of $\xi$ and $\theta$ are shown in Fig.~\ref{fig:A}, where  $1\sigma$ errors of data in BRs with units of $10^{-3}$
 are taken. From this figure, we constrain the free parameters as 
 \begin{eqnarray}
  -5.3\times 10^{-2} <\xi <-3.\times 10^{-2},\;\;  1.47<\theta< 1.87. 
 \end{eqnarray}
 
\begin{figure}[htbp]
\includegraphics*[width=5 in]{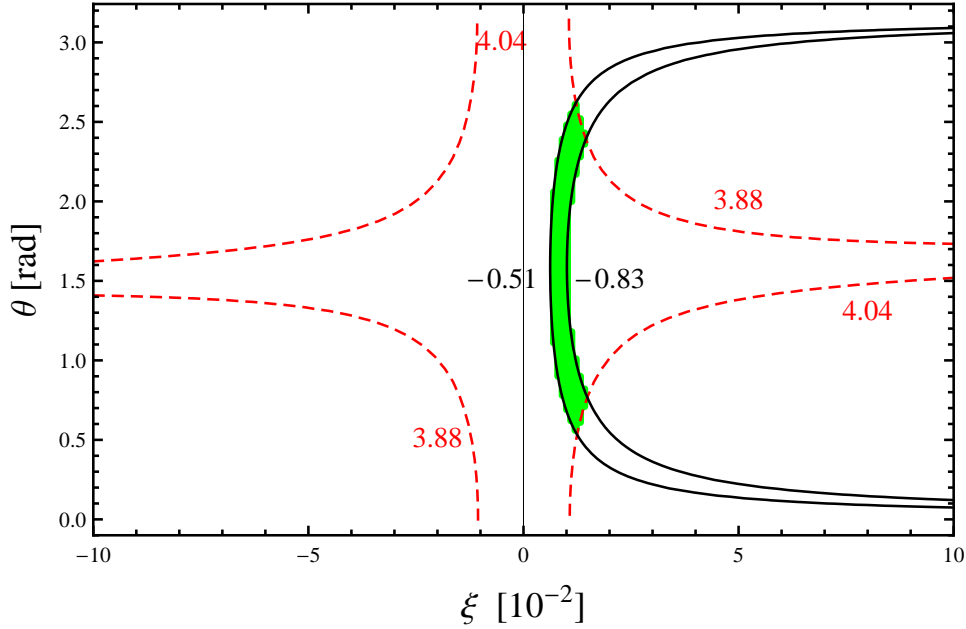}
\caption{BR for  $D^0\to K^+ K^-$ (dash-dotted) and $\Delta A_{CP}$ (solid), where the shaded band stands for the allowed region.  }
 \label{fig:B1}
\end{figure}

\underline{\it{Case} II:} 
In this case, Eq.~(\ref{eq:damp2}) becomes
 \be
|A^s|^2 -|\bar A^s|^2 &=& -4 E^s_{SM}T^s_{LR}\sin\delta^s_S  \frac{a_1'}{a_1} \xi   \lambda^2\left(  Im V^{R*}_{ud} +  Im V^R_{cs} \right)\,. \label{eq:damp2-2}
 \ed
Without further limiting  the pattern of $V^R$, apparently the situation in Case II is more complicated. 
It was pointed out that one can have a weaker constraint on the mass of $W_R$  when the right-handed flavor mixing matrix is centered around 
the following two forms \cite{Langacker:1989xa}:
 \be
V^R_{A}(\alpha) =
  \left( \begin{array}{ccc}
   1 & 0 & 0 \\ 
    0 & c_\alpha & \pm s_\alpha \\ 
    0 & s_\alpha & \mp c_\alpha \\
  \end{array} \right)\,, \ \ \  V^R_{B}(\alpha) =
  \left( \begin{array}{ccc}
   0 & 1 & 0 \\ 
    c_\alpha & 0 & \pm s_\alpha \\ 
    s_\alpha &  0 & \mp c_\alpha \\
  \end{array} \right)\,,
 \ed
where $c_\alpha=\cos\alpha$, $s_\alpha=\sin\alpha$ and $\alpha$ is  an arbitrary angle. We note that the null elements denote the values that are smaller than $O(\lambda^2)$, thus their effects could be ignored in the analysis. We will focus on the implication of the two special forms.
 In $V^R_A(\alpha)$, since $V^R_{cd}\to 0$ and $Im(V^R_{ud})\to 0$  due to $(\epsilon'/\epsilon)_K$, 
  only the CPA for $D^0\to K^+ K^-$ could be compatible with the current data, while the CPA for $D^0 \to \pi^+ \pi^-$ decay is small, i.e. $\Delta A_{CP}\approx A_{CP}(D^0\to K^+ K^-)$. With $\alpha\approx 0$ which satisfies the constraint from $b\to d \gamma$ \cite{Crivellin:2011ba}, we present the constraint of ${\cal B}(D^0\to K^+ K^-)$ and 
 $\Delta A_{CP}$ as functions of  $\bar \xi=g_R/g_L \xi$ and $\theta=$arg$(\bar V^R_{cs})$ in Fig.~\ref{fig:B1}, where 
 the shaded band  shows the allowed region for the parameters, corresponding to 
 \begin{eqnarray}
  0.7\times 10^{-2} <\xi <1.4\times 10^{-2},\;\;  0.56<\theta< 2.61. 
 \end{eqnarray}

\begin{figure}[htbp]
\includegraphics*[width=5 in]{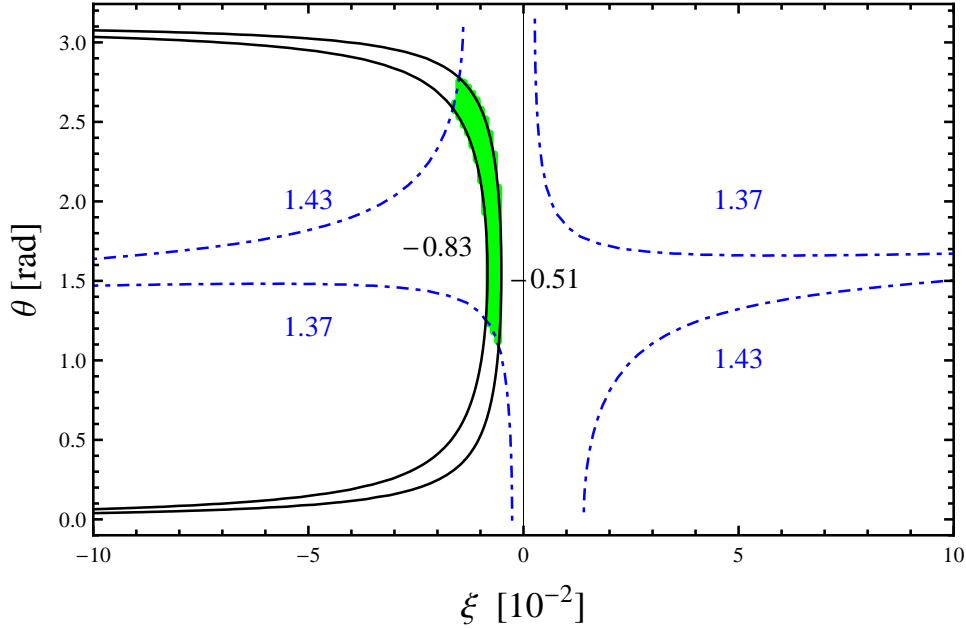}
\caption{The Legend is the same as Fig.~\ref{fig:B1} but for $D^0\to \pi^+ \pi^-$.}
 \label{fig:B2}
\end{figure}

For $V^R_B(\alpha)$,  due to $V^R_{ud, cs}\to 0$, the CPA for $D^0\to K^+ K^-$ is small and only $A_{CP}(D^0\to \pi^+ \pi^-)$ 
could be compatible with the data. As a result, we have $\Delta A_{CP}\approx - A_{CP}(D^0\to \pi^+ \pi^-)$. Similar to $V^R_A(0)$ with $\alpha=0$, 
we display   ${\cal B}(D^0\to \pi^+ \pi^-)$ (dashed) and  $\Delta A_{CP}$ (solid) as functions of $\bar\xi$ and the phase $\theta$  defined as $\bar V^R_{cd}=-\lambda e^{-i\theta}$ in Fig.~\ref{fig:B2}. In this case, the allowed $\xi$ is negative
 \begin{eqnarray}
  -1.6\times 10^{-2} <\xi <-0.6\times 10^{-2},\;\;  1.12<\theta< 2.76. 
 \end{eqnarray}

In summary, we have studied the impact of the LR mixing in the general LR model on the CPA difference between $D^0\to K+ K^-$ and $D^0\to \pi^+ \pi^-$.  It is found that when the constraint from  $(\epsilon'/\epsilon)_K$ is considered, the proposed LR mixing mechanism could be compatible with the value of $\Delta A_{CP}$  averaged by the LHCb and CDF new data.  To illustrate the influence of the LR mixing effects, we have adopted two cases for the new flavor mixing matrix $\bar V^R$
to explain the large $\Delta A_{CP}$. 
In Case I, we have found that $A_{CP}(D^0\to K^+ K^-)\approx - A_{CP}(D^0\to \pi^+ \pi^-)$ can be achieved. 
In Case II, we have used two special forms for $V^R$, resulting in  $A_{CP}(D^0 \to \pi^+ \pi^-) \approx 0$  and  $A_{CP}(D^0\to K^+ K-)\approx 0$, respectively.\\

 \noindent {\bf Acknowledgments}

This work is supported by the National Science Council of R.O.C.
under Grant \#s: NSC-100-2112-M-006-014-MY3 (CHC) and
NSC-98-2112-M-007-008-MY3 (CQG).

\end{document}